\documentclass[4pt,twocolumn]{article}
\pdfoutput=1
\usepackage[lmargin=0.75in,rmargin=0.75in,tmargin=1.0in,bmargin=1.0in]{geometry}
\usepackage{wrapfig}

\usepackage{graphicx}
\usepackage{pdfpages}
\usepackage[switch,columnwise]{lineno}

\usepackage[font={small}]{caption} 

\usepackage[square,numbers,sort]{natbib}
\usepackage{authblk}

\usepackage{hyperref} 
\usepackage{chngcntr}

\begin{document}
\title{Discovering How Ice Crystals Grow Using Neural ODE's and Symbolic Regression}
\date{}
\setcounter{Maxaffil}{1}

\author[a,b,*]{Kara D. Lamb}
\author[c]{Jerry Y. Harrington}
\author[c]{Alfred M. Moyle}
\author[d]{Gwenore F. Pokrifka}
\author[e]{Benjamin W. Clouser}
\author[f,g,h]{Volker Ebert}
\author[i]{Ottmar M\"{o}hler}
\author[i]{Harald Saathoff}

\affil[a]{Department of Earth and Environmental Engineering, New York, NY, 10027, USA}
\affil[b]{Learning the Earth with Artificial Intelligence and Physics Center, Columbia University, New York, NY, 10027, USA}
\affil[c]{Department of Meteorology, Pennsylvania State University, State College, PA, 16802, USA}
\affil[d]{High Meadows Environmental Institute, Princeton University, NJ, 08540, USA}
\affil[e]{Department of the Geophysical Sciences, University of Chicago, Chicago, IL, 60637, USA}
\affil[f]{Physikalisch Chemisches Institut, Heidelberg University, INF 253, 69116 Heidelberg, Germany}
\affil[g]{Physikalisch-Technische Bundesanstalt, Bundesallee 100, 38116 Braunschweig, Germany}
\affil[h]{Reactive Flows and Diagnostics, Technische Universit\"{a}t Darmstadt, Darmstadt, Germany}
\affil[i]{Institute of Meteorology and Climate Research Atmospheric Aerosol Research, Karlsruhe Institute of Technology, 76344 Eggenstein-Leopoldshafen, Germany}
\affil[*]{Corresponding author (kl3231@columbia.edu)}

\maketitle

\begin{abstract} 
\textbf{Depositional ice growth is an important process for cirrus cloud evolution, but the physics of ice growth in atmospheric conditions is still poorly understood. One major challenge in constraining depositional ice growth models against observations is that the early growth rates of ice crystals cannot be directly observed, and proposed models require assumptions about the functional dependence of physical processes that are still highly uncertain. Neural ordinary differential equations (NODE's) are a recently developed machine learning method that can be used to learn the derivative of an unknown function. Here we use NODE's to learn the functional dependence of unknown physics in the depositional ice growth model by optimizing against experimental measurements of ice crystal mass. We find a functional form for the depositional ice growth model that best fits 290 mass time series of ice crystals grown in a levitation diffusion chamber. We use symbolic regression to derive an equation for the function learned by the NODE model, which includes additional terms proportional to ice crystal mass in the capacitance growth model. We evaluate this functional form against experimental data sets from the AIDA Aerosol and Cloud Chamber, finding that our new proposed model for depositional ice growth accurately reproduces experimental results in the early stages of ice crystal growth.}
\end{abstract}

Cirrus clouds, pure ice clouds that form in the Earth's upper troposphere, play a significant role in the climate cycle \cite[]{gasparini_opinion_2023}. Mid-latitude and tropical cirrus cover approximately one third of the upper troposphere at any point in time, and have significant direct radiative climate effects. In addition to their radiative effects, the high, thin cirrus clouds that form in the Earth's tropical tropopause layer act as an important control on the amount of water vapor that enters the stratosphere from the troposphere, where it is a significant greenhouse gas \cite[]{solomon_contributions_2010}. Contrail cirrus forming as the result of commercial aviation represent a significant, but uncertain, anthropogenic climate forcing \cite[]{karcher_formation_2018}. Uncertainties in ice microphysical process rates that control cirrus cloud formation and lifetime translate to direct climate radiative forcing uncertainties on the order of 30 W/m$^2$, more than 8 times the radiative forcing associated with a doubling of CO$_{2}$ in the atmosphere \cite[]{sullivan2021ice,lamraoui2023sensitivity}. 

\begin{figure*}
\begin{center}
\noindent\includegraphics[width=\textwidth]{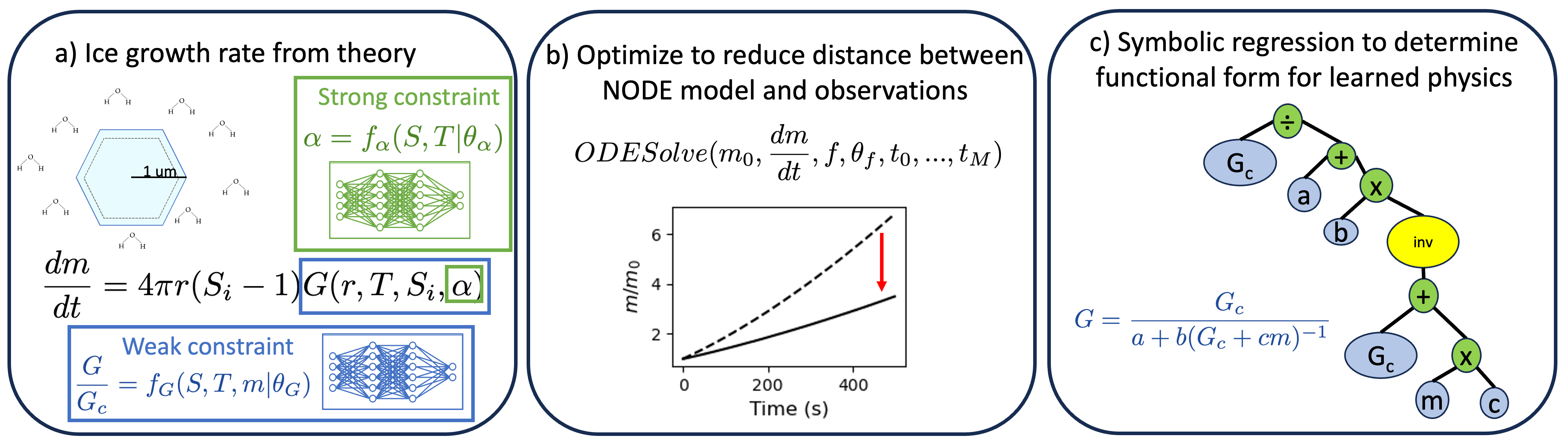}
\caption{\textbf{Overview of methodology for learning unknown physics in depositional ice growth models.} Depositional ice growth is an important process for ice formation in atmospheric clouds. Ice crystals grow via direct deposition of water molecules from the vapor phase onto the ice surface. a) We replace partially unknown physics in the depositional ice growth model with a neural network, considering both a strong and weak constraint. b) We integrate the ice growth rate partially parameterized by a neural network and optimize to reduce the distance between the model and experimentally measured time series of ice mass ratios. c) We use symbolic regression to determine a functional form for the unknown physics learned by the neural network.}\label{fig:overviewmethods}
\end{center}
\vspace{-5mm}
\end{figure*}

One of the most important microphysical processes controlling the growth and evolution of cirrus clouds is the depositional ice growth rate, which describes the adsorption, surface migration, and incorporation of water molecules into the ice crystal (Figure \ref{fig:overviewmethods}a). 
However, significant gaps in our understanding of depositional ice growth, particularly the early growth of ice crystals in clouds, continue to limit our ability to accurately model these processes \cite[]{avramov2010influence,sullivan2021ice}. Early growth is especially hard to constrain because this growth connects freshly nucleated particles to larger faceted crystals. These small ice crystals (radii $<$ 50 $\mu$m) undergo various surface transformations \citep[]{gonda1978morphology,gonda1984initial, pokrifka2020estimating, rui2025growth} as habit forms develop, affecting the growth rates.

Structural and parametric uncertainty in ice growth models has been difficult to address \cite[]{morrison2020confronting}. Structural uncertainty refers to uncertainty in the functional dependence of a physical model, while parametric uncertainty refers to uncertainty in parameter values. While depositional ice growth is typically represented in cloud and climate models with capacitance theory \cite[]{pruppacher1998microphysics}, the early stages of ice growth leading to the formation and growth of facets is determined by surface attachment kinetics, which are poorly constrained by current theories, observations (those made in clouds), and laboratory measurements. Measurements of surface attachment kinetics modeled as a deposition coefficient from different experimental studies disagree by orders of magnitude \cite[]{magee2006experimental, skrotzki2013accommodation}, although recent work has suggested these studies can be reconciled with a saturation and temperature dependent functional form \cite[]{harrington2019calculating,lamb2023re}. One significant challenge in constraining depositional ice growth rates in laboratory studies is that the early stages of growth cannot be directly observed, and must instead be inferred from observables such as the mass, and often in conditions where one or more dependent variables (such as temperature or supersaturation with respect to ice) are held constant.

Here we use physics-informed machine learning to address structural uncertainty in the depositional ice growth model. Scientific machine learning for discovering unknown physics directly from observations has demonstrated significant promise in recent years \cite[]{schmidt2009distilling,cranmer2020discovering,udrescu2020ai}. Approaches such as physics-informed neural networks (PINN's) \cite[]{raissi2019physics} can be leveraged to integrate observational data with known governing physical laws, even in cases with partially unknown physics. Neural ordinary differential equations (NODE's) integrate neural networks parameterizing an unknown function with typical numerical ODE solvers to learn a continuous model for a physical system \cite[]{chen2018neural}. In this study, we develop a methodology to learn the functional dependence for the ODE describing the depositional ice growth rate by optimizing a NODE across multiple time series simultaneously. We compare strongly-constrained and weakly-constrained NODE models, based on the amount of prior physical knowledge that is included. We apply this method to experimental measurements to learn a mathematical model for the depositional ice growth rate and use symbolic regression to discover a closed form expression for the non-linear relationship learned by the neural network. Finally, we evaluate this new proposed model on an independent data set of cirrus cloud experiments performed in the AIDA Aerosol and Cloud Chamber.

\section*{Levitation Diffusion Chamber Observations}
Levitation diffusion chamber experiments provide time-resolved measurements of individual ice crystals growing in a highly-controlled environment. Here we explore how scientific machine learning and equation discovery can be applied to measurements of both heterogeneously and homogeneously-nucleated ice crystals grown in a levitation diffusion chamber at temperatures between 205 - 240 K and saturation with respect to ice between 1.0 and 1.8, conditions characteristic of upper tropospheric cirrus cloud formation \cite[]{harrison2016levitation,pokrifka2020estimating,pokrifka2023effective}. During experiments, individual ice crystals with initial radii between 6 and 26 $\mu$m are nucleated, levitated, and grown from vapor in constant saturation $S_{i}$ and temperature $T$ conditions. Measurements consist of 290 time series of the mass ratio ($m/m_{0}$) of individual ice crystals, where $m$ is the mass of an ice crystal, and $m_{0}$ is its initial mass. The crystals remain relatively small over the course of the experiment, with equivalent spherical radii typically $<$ 40 $\mu$m (max. of $\sim$84 $\mu$m). Since mass ratio time series have varying durations (depending on how efficiently the ice crystals grow), we interpolate time series to 1 Hz and crop all data sets to a maximum length of 500 seconds. The $S_{i}$, $T$, and mass time series for all 290 experiments used in the analysis are shown in Supplementary Figure S1.

In addition to the experimentally measured ice crystal mass ratio time series, we create a synthetic data set of levitation diffusion chamber measurements with a known functional dependence for the depositional growth rate to validate the performance of our equation discovery method. This synthetic data set consists of 290 mass ratios of ice crystals growing in the same saturation $S_{i}$ and temperature $T$ conditions and with the same initial mass $m_{0}$, but with an assumed functional form for the depositional ice growth model based on Nelson and Baker, 1996 \cite[]{nelson1996new} (Supplementary Figure S2). Realistic measurement noise is added to the synthetic time series by using the trailing eigenmodes of a singular spectrum analysis decomposition applied to the measured time series (Supplementary Figure S3 shows an example). Additional details about experimental and synthetic data sets are given in \textbf{\nameref*{sec:matmethods}}. 

\section*{Methodology to Learn Unknown Physics}

In this section, we describe the scientific machine learning methodology we use to discover a mathematical model for depositional ice growth that best fits all 290 levitation diffusion chamber experiments. An overview of the method is shown in Figure \ref{fig:overviewmethods}. The depositional ice growth rate for an ice crystal growing from vapor has traditionally been modeled using the capacitance ice growth model \cite{pruppacher1998microphysics}, which is modified to include surface kinetics for the early stages of ice growth\cite[]{harrington2019calculating}. This model is an ODE,
\begin{equation}\label{eq:dmidt_simplified}
\frac{dm}{dt} = 4\pi r(S_{i}-1)G(r,T,S_{i},\alpha),
\end{equation}
where $m$ is the mass of the crystal, $r$ is the radius of the ice crystal, and $S_{i}$ is the ambient ice supersaturation. The function $G(r,T,S_{i},\alpha)$ represents the combined effects of gas-phase vapor and thermal diffusion along with surface attachment kinetics typically parameterized by a deposition coefficient $\alpha$. The deposition coefficient $\alpha$ has previously been parameterized as a saturation and temperature-dependent function in Nelson and Baker, 1996 (Eq.~\ref{eq:AlphaNelson}) \cite[]{nelson1996new}. However, this form may not be valid for crystals undergoing rapid transformations immediately following nucleation \citep[]{gonda1984initial}. A more detailed description of the capacitance ice growth model with modifications for surface kinetics is provided in \textbf{\nameref*{sec:matmethods}}.

Predicting ice growth in the levitation diffusion chamber amounts to solving an initial value problem, where the ice crystal mass as a function of time is given by
\begin{equation}\label{eq:initialvalue}
m(t) = \int_{0}^{t}\frac{dm}{dt}dt+m_{0}.
\end{equation}
If we knew the functional dependence of Eq.~\ref{eq:dmidt_simplified}, then this problem would be straightforward to solve. However, the functional dependence for $G(r,T,S_{i},\alpha)$ is unknown. This uncertainty arises from the difficulty in consistently parameterizing various surface transformations ice crystals undergo in their early stages of growth, and the lack of observational constraints on the evolving habit of these micron-scale ice crystals. While prior attempts to parameterize size-based surface kinetics due to transformations have been made, most have been either ad hoc \citep{Giere:2003} or lack generality due to the use of smaller data sets \citep{Harrington:2021a}. In addition, since each ice crystal grows at a constant, but unique $S_{i}$ and $T$, determining a functional dependence of $G(r,T,S_{i},\alpha)$ on $r$, $T$, $S_{i}$, and with a possibly saturation and temperature-dependent $\alpha$ that is consistent across all experiments is not straight-forward. 

Here we focus on learning a single depositional ice growth model that best fits all experiments, with the aim of developing a simple parameterization for cirrus cloud models. We use a NODE model to solve Eq~\ref{eq:initialvalue} to learn a consistent functional form for $G(r,T,S_{i},\alpha)$ across all 290 experiments simultaneously. We assume two cases for prior physical knowledge in Eq.~\ref{eq:dmidt_simplified} based on past literature (Figure \ref{fig:overviewmethods}a). The first model makes a stronger assumption about the amount of prior physical knowledge to include, while the second model learns a greater part of the ice growth model from experimental measurements. In the first case, we use a capacitance model for ice growth that assumes an unknown function for the deposition coefficient $\alpha$, which we refer to as the ``strongly-constrained NODE" model. In the second case, we fit the ratio of the transfer coefficient $G$ relative to the theoretical transfer coefficient $G_{c}$ for a spherical ice crystal assuming the continuum limit (Eq.~\ref{eq:Gcont}), which we refer to as the ``weakly-constrained NODE" model. In each case, we use prior physical knowledge from Eq.~\ref{eq:dmidt_simplified} and replace only the uncertain portion of the model with a neural network. For the strongly-constrained NODE model, we assume the functional form for $G(r,T,S_{i},\alpha)$ is given by Eq.~\ref{eq:g_sk}, and we further assume $\alpha$ is a function of temperature and supersaturation,
\begin{equation}\label{eq:alphaNN}
\alpha = f_{\alpha}(S_{i},T|\theta_{\alpha}).
\end{equation}
The expression $f_{\alpha}(S_{i},T|\theta_{\alpha})$ represents a neural network that takes as input $S_{i}$ and $T$ and given the weights of the neural network $\theta_{\alpha}$ predicts a value for $\alpha$. For the weakly constrained NODE model, we assume the ratio between $G$ and $G_{c}$ is given by a function of the temperature, supersaturation, and mass of the ice crystal,
\begin{equation}\label{eq:GGcNN}
\frac{G}{G_{c}} = f_{G}(S_{i},T,m|\theta_{G}).
\end{equation}
Parameterizing kinetics with a modified $G$ is advantageous because it is more general, and it has some measurement \citep[]{pokrifka2020estimating} and theoretical \citep[]{harrington2009parameterization} backing. 

To compare against experimental measurements (Figure \ref{fig:overviewmethods}b), we integrate Eq. \ref{eq:dmidt_simplified} partially parameterized by neural networks using an ODE solver (implemented in \texttt{PyTorch} with the \texttt{torchdiffeq} library \cite[]{chen2018neural}). We optimize the strongly and weakly constrained NODE models against the experimental time series to simultaneously learn the neural network weights $\theta_{\alpha}$ and $\theta_{G}$ in Eq.~\ref{eq:alphaNN} and Eq.~\ref{eq:GGcNN}, respectively, that reduce the distance between the models and the measured time series. Since we assume all 290 experimental mass time series can be modeled with the same physical model, we optimize the NODE models across all time series simultaneously by minimizing L2 loss (or mean square error, MSE, loss) between the measured time series and the model, 
\begin{equation}\label{eq:MSEloss}
\mathcal{L_{NODE}} = \sum_{j=0}^{N_{exp}}\sum_{i=0}^{T_{j}} \left(\frac{m_{j}(t)}{m_{j,0}}-\frac{\hat{m}_{j}(t)}{m_{j,0}}\right)^{2},
\end{equation}
where $j$ is the experiment number, and $T_{j}$ is the length of the $j$th time series, and $\hat{m}_{j}(t)$ indicates the prediction from the NODE models, and $N_{exp}=290$. 

Once the NODE models have been optimized against experimental measurements, we use symbolic regression to derive functional forms for the trained neural networks (Figure \ref{fig:overviewmethods}c). Symbolic regression uses genetic algorithms to search for equations that optimize accuracy while limiting model complexity. Here we use the symbolic regression library \texttt{PySR} to return a Pareto front of candidate expressions \cite[]{cranmer2023interpretable}. We use the observed values for $S_{i}$ and $T$, and the initial mass $m_{0}$ for each of the 290 experiments. These values are used as input for the trained neural networks to determine the corresponding values for $\alpha$ for the strongly-constrained NODE model (Eq. \ref{eq:alphaNN}) and for $G$ for the weakly-constrained NODE model (Eq. \ref{eq:GGcNN}). Further details of the strongly and weakly-constrained NODE models, the training process, and the equation discovery method are given in \textbf{\nameref*{sec:matmethods}}. 

\section*{Results}
\begin{figure*}
\noindent\includegraphics[width=\textwidth]{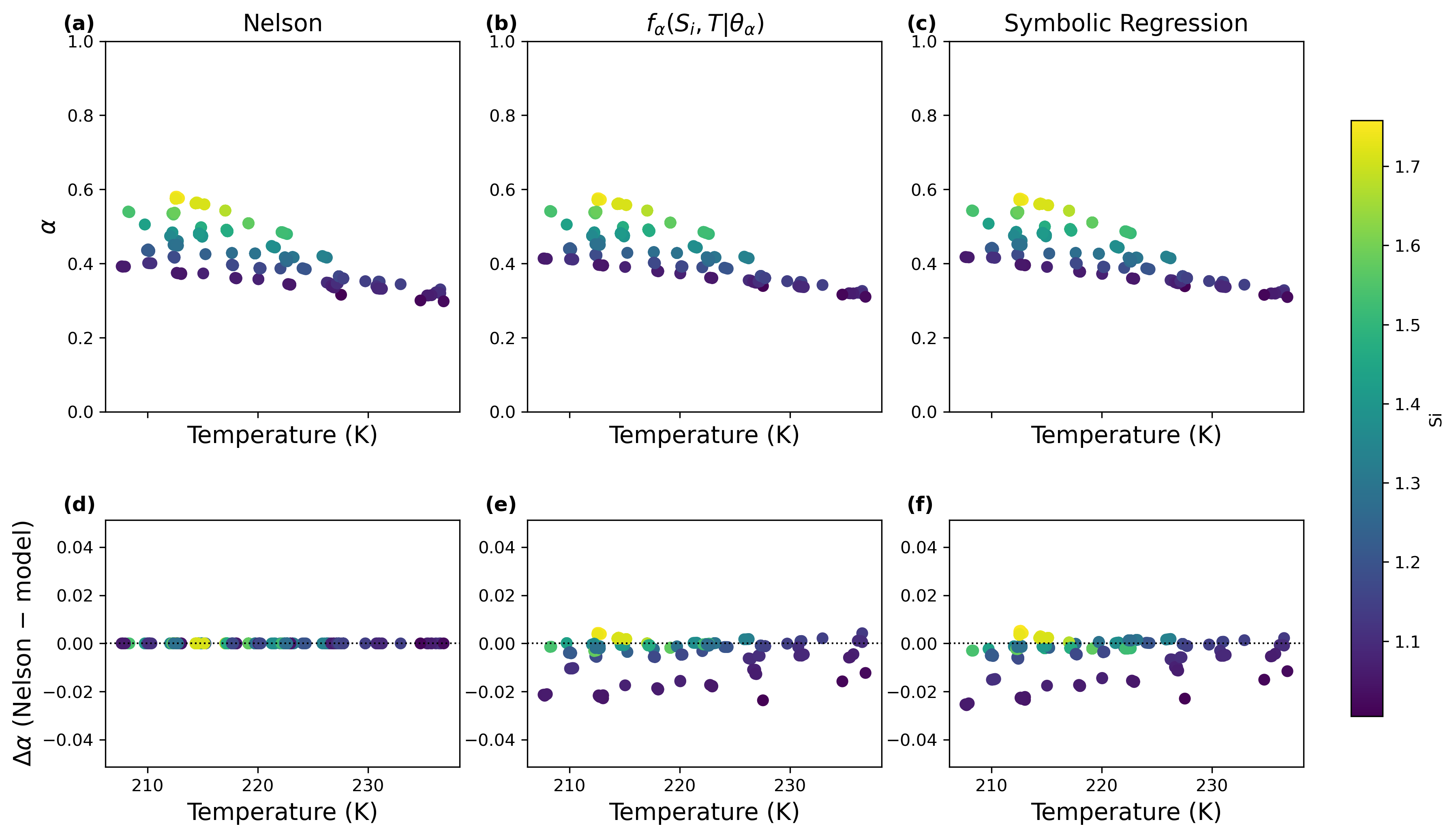}
\caption{\textbf{Functional dependence of $\alpha$ learned from synthetic data sets.} a) Saturation and temperature dependence of $\alpha$ parameterization from Nelson and Baker, 1996 \cite[]{nelson1996new}. b) Saturation and temperature dependence from the trained NN for the synthetic data sets. c) Predictions from an expression learned by symbolic regression from the trained NN for the synthetic data sets. d - f) Residuals between $\alpha$ from Nelson and Baker, 1996 and model predictions for $\alpha$.}
\label{fig:synthdata_recoveredalpha}
\end{figure*}

We first test our method on the synthetic data sets with a known functional form for the deposition coefficient $\alpha$ based on \cite[]{nelson1996new}. After optimizing the strongly-constrained NODE model against the synthetic data sets, we find that the model is able to reproduce the time series very accurately (Supplementary Figure S4 shows examples of a subset of time series compared to NODE model fits). We compare the predicted values of $\alpha$ from the trained neural network $\alpha_{\textrm{NN}}$ to those used to generate the synthetic data set $\alpha_{\textrm{Nelson}}$. The strongly-constrained NODE model is able to learn the non-linear functional dependence for $\alpha$ (Eq. \ref{eq:AlphaNelson}) that closely matches the one used to generate the synthetic data sets (Figure \ref{fig:synthdata_recoveredalpha}, left and middle panels). In addition, when we use symbolic regression to learn a functional dependence for $\alpha$ from the trained neural network, it closely matches the true dependence (Figure \ref{fig:synthdata_recoveredalpha}, right panel).

We next use our approach on the experimentally measured time series. Since we have no ground-truth for the depositional ice growth model for the experimental data sets, we compare the performance of the weakly and strongly-constrained NODE models optimized against the experimental data sets with the model from Nelson and Baker, 1996 \cite[]{nelson1996new} and a model with no surface kinetics (Table \ref{table:modelperformance}). We evaluate the MSE loss between models and experimental measurements across all 290 experiments (Eq. \ref{eq:MSEloss}), as well as evaluating which model performs best on the majority of individual experiments (determined by the model which gives the lowest MSE loss for an individual experiment). According to both metrics, the weakly-constrained NODE model outperforms a model with no surface kinetics, the Nelson and Baker, 1996 model and the strongly-constrained NODE model. The weakly-constrained NODE model performs best on 138 out of 290 experiments (Table \ref{table:modelperformance}, 3rd column). When considering performance on individual experiments, the strongly-constrained NODE model performs better than the Nelson and Baker, 1996 parameterization, but only slightly better than a model with no surface kinetics.

\begin{table*}
\caption{\textbf{Comparison of model performance on experimental data from the levitation diffusion chamber.} MSE loss between model and experiments is evaluated across all 290 time series; we separately evaluate the interpolation performance (t$<$500 s) and the extrapolation performance (t$<$1000s). Best experiment count refers to the total number of individual experiments for which the proposed model performed best.\label{table:modelperformance}}
\centering

\begin{tabular}{lcccc}
\hline
& \textbf{MSE Loss} & \textbf{MSE Loss} & \textbf{Best Exp. Count} & \textbf{Best Exp. Count}\\
\textbf{Ice growth model}  & \textbf{(t$<$500 s)} & \textbf{(t$<$1000 s)} & \textbf{(t$<$500 s)} & \textbf{(t$<$1000 s)}\\
\hline
No surface kinetics ($G$ = $G_{c}$)& 41396 & 222750  & 62 & 55 \\
Nelson and Baker, 1996 \cite[]{nelson1996new} & 40751 & 220930 & 29 & 31 \\
Strongly-constrained NODE model & 30714 & 203645 & 61 & 62  \\
Weakly-constrained NODE model & \textbf{16495} & \textbf{184723} & \textbf{138} & \textbf{142}\\

\hline
\end{tabular}
\end{table*}

\begin{figure*}
\begin{center}

\noindent\includegraphics[width=1.0\textwidth]{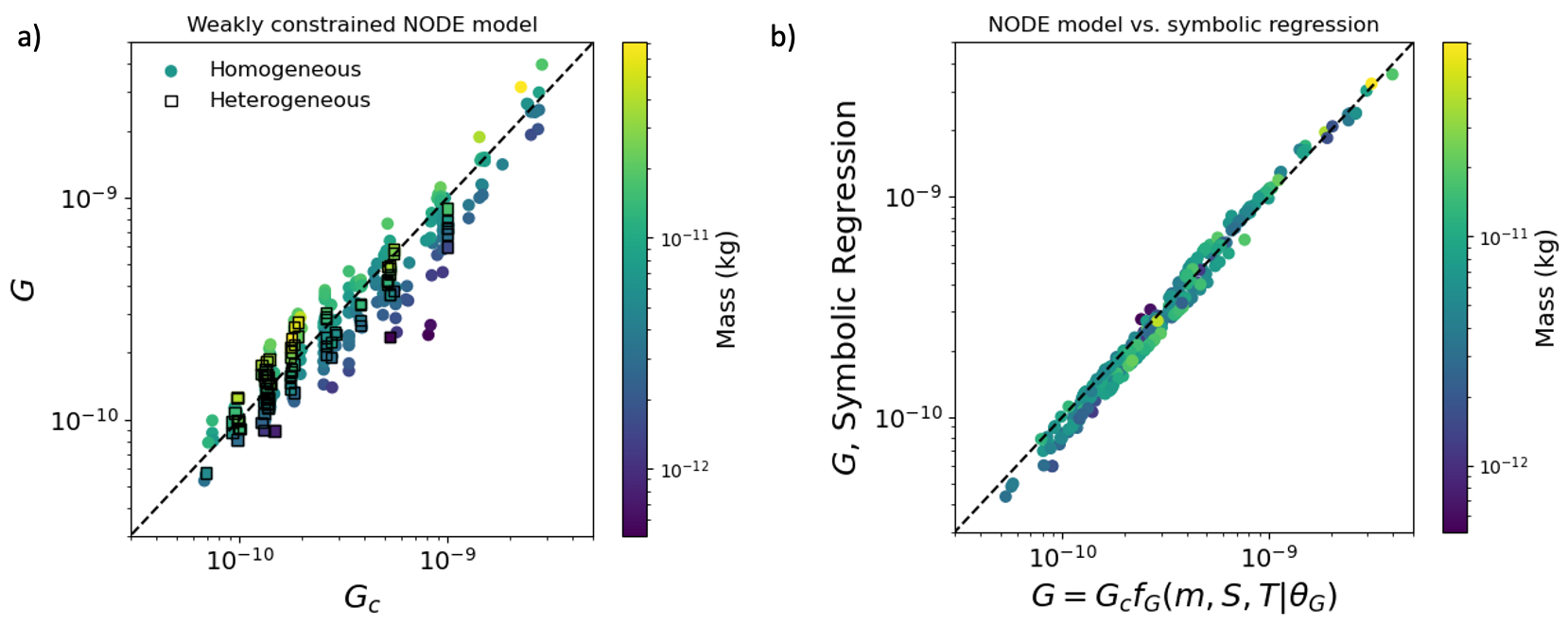}
\caption{\textbf{Unknown physics learned by weakly-constrained NODE model from experiments.} a) The transfer coefficient $G$ learned by the weakly-constrained NODE model for the 290 experiments compared with $G_{c}$ (the transfer coefficient for the continuum case, assuming a spherical ice crystal). b) Predictions from the trained neural network for $G$ compared with the predictions from an expression learned by symbolic regression from the trained neural networks (Eq. 8 in Table S2) for the 290 experiments.}
\label{fig:learnedphysicsG}
\end{center}
\end{figure*}

In comparing the strongly-constrained NODE model optimized against the experimental data sets with the measured mass ratios, we find significant deviations between the measured mass ratios and predicted mass ratios for some experiments (Supplementary Figure S5). The predicted values of $\alpha$ from the trained neural network also differ significantly from those predicted by the functional form given in \cite[]{nelson1996new}, and suggest that the neural network struggles to find a consistent smoothly varying function across $S_{i}$ and $T$ (Supplementary Figure S6). As the strongly-constrained NODE model is not able to learn a functional dependence for $\alpha$ that reproduces the mass ratio time series for the majority of the experiments, this suggests that a model using a single deposition coefficient form $\alpha$ may not be the optimal functional relationship to describe the depositional ice growth process for newly formed crystals. By contrast, when we optimize the weakly-constrained NODE model against the experimental data sets, this generally reduces deviations between the mass ratio time series and the predicted values for the mass ratios (Supplementary Figure S5) compared to the strongly-constrained NODE model. In addition, we can use the trained neural network to evaluate the learned functional dependence of $G$ compared with $G_{c}$ (Figure \ref{fig:learnedphysicsG}a), and our model learns a consistent functional dependence across the 290 experiments. While $G$ is generally within a factor of 1.2 of $G_{c}$ for most experiments, the learned functional dependence indicates there is an additional dependence on $m$ that is not accounted for by $G_{c}$. 

As described in the previous section and in \textbf{\nameref*{sec:matmethods}}, we use symbolic regression to determine a functional dependence for $G$ learned by the neural network. Symbolic regression discovers multiple candidate expressions for G (See Table S2). All discovered expressions depend on $G_{c}$ (as given by given by Eq. \ref{eq:Gcont}) and $m$, the crystal mass.  Only the most complex expressions show any dependence on $S_{i}$ and $T$, indicating that the majority of the variance in the physical model can be related to the ice crystal size.  This functional dependence is consistent with physical expectations for surface kinetic effects, which should suppress ice growth when ice crystals are small and habits are just forming. More complex expressions typically have lower MSE loss (Figure S7), and we further validate these expressions on independent data (discussed in the next section) to determine which equation provides the optimal balance between complexity and generalization performance.
\subsection*{Performance of proposed models on unseen data}
To further validate the learned depositional ice growth models, we evaluate our new proposed models for depositional ice growth on data that was unseen during the training process of the NODE model. 

Since we only used the initial 500 seconds from each time series for the ice crystals grown in the levitation diffusion chamber, we first test how well models perform on predicting the mass ratio of the ice crystals for the remainder of the experimental time series (e.g. for t$>$500 s). However, we do note that this extrapolation test will necessarily be biased towards ice crystals that are compact and grow less efficiently, typically at lower supersaturations. High supersaturation promotes polycrystalline growth, which is rapid, and the time series measurements for such crystals are relatively short, and thus may not be represented in this data set. We test how well the models perform by integrating the time series up to 1000 s and evaluating the MSE loss between the measured time series and the model (Table \ref{table:modelperformance}, 2nd column). In this case, the weakly-constrained NODE model still performs best, but the performance is more comparable with the other three proposed models. In addition, when integrating for significantly longer times greater than 1000 s, the weakly-constrained NODE model begins to perform worse relative to the other three models, as the weakly-constrained NODE model begins to over-predict ice growth at larger sizes, which is consistent with longer time series measurements being associated with less efficient growth. This suggests that the weakly-constrained NODE model is an improvement on the previous models when $m$ is small, but that in the limit where $m$ is larger, depositional ice growth should asymptote to the continuum limit. 

\begin{figure}
\begin{center}
\noindent\includegraphics[width=0.45\textwidth]{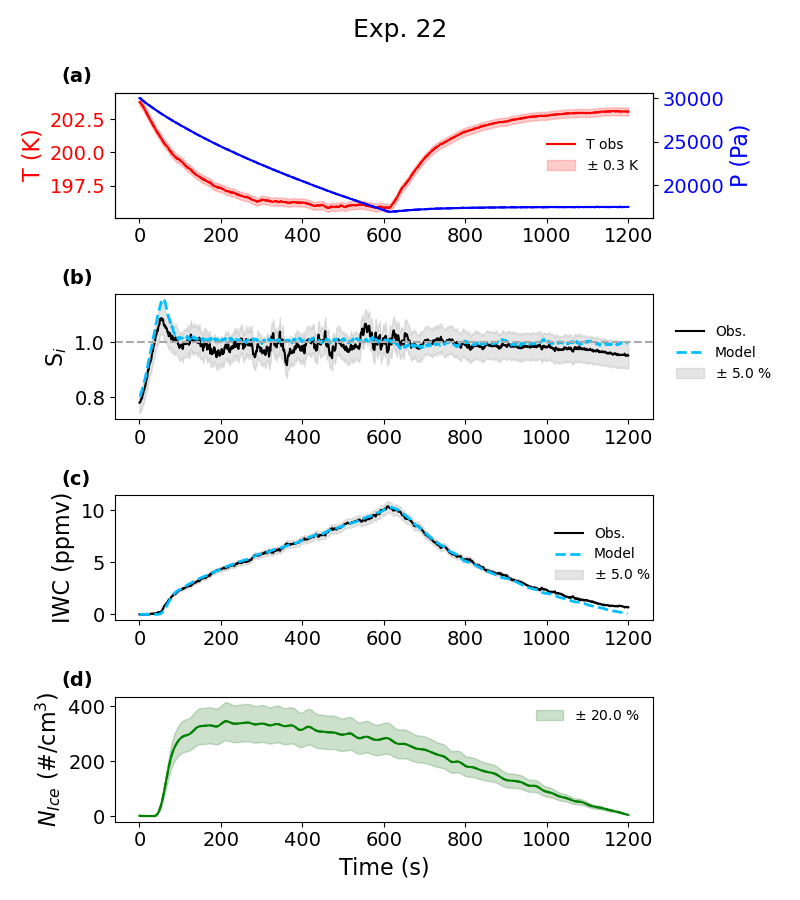}
\caption{\textbf{Comparison of learned depositional ice growth model to a cold cirrus cloud experiment in the AIDA Cloud Chamber.} a) Temperature and pressure during the adiabatic expansion experiment. b) Observed supersaturation with respect to ice inside the chamber,  compared with predicted supersaturation with respect to ice. Gray dashed line indicates supersaturation equal to 1.0. c) Observed ice water content compared to predicted ice water content. d) Observed ice number density inside of the chamber.}
\label{fig:exp22}
\end{center}
\end{figure}

Next, we evaluate how well the proposed depositional ice growth models perform on a completely independent data set, using observations from adiabatic expansion experiments simulating cirrus cloud formation in the AIDA Aerosol and Cloud Chamber during the IsoCloud campaign \cite[]{lamb2017laboratory,clouser_no_2020,lamb2023re}. These experiments heterogenously and homogenously nucleated populations of ice crystals, which then grow in conditions characteristic of upper tropospheric cirrus cloud formation over approximately 10 minutes. The $S_{i}$ and $T$ range for the AIDA experiments overlap with the levitation diffusion chamber experiments but also include experiments at lower T and higher $S_{i}$, and time-evolving (rather than fixed) environmental conditions (See Figure S8). In AIDA we observe the total mass of the population of ice crystals in a volume of air (the ice water content, IWC) as a function of time, rather than the mass of individual ice crystals. In order to compare our proposed depositional ice growth models against the AIDA experiments, we model ice growth in AIDA using a parcel model with bin microphysics constrained to the observed number concentrations of ice, as was previously described in \cite[]{lamb2023re}. We replace the depositional ice growth model with the learned expressions from symbolic regression for $G$ given in Table S2, and use the AIDA experiments to explore which proposed functional form for $G$ best fits the majority of AIDA experiments. We find that the following functional form (Discovered Equation 8 in Table S2),
\begin{equation}\label{eq:newG}
    G =  a_{0} G_{c}^{a_{1}}\left[a_{2} + \frac{a_{3}}{m}\right]^{-1}+a_{4} , 
\end{equation}
where $a_{0} = 688.267$, $a_{1} = 1.3153$, $a_{2} = 0.85601$, $a_{3} = 2.6606\times10^{-12}$, and $a_{4} = 0.1123\times10^{-9}$, improves upon the capacitance ice growth (e.g. no surface kinetics) and the Nelson and Baker, 1996 models at lower temperatures in AIDA (Figure S9). Figure \ref{fig:exp22} shows an example AIDA experiment modeled with the new proposed model for depositional ice growth using Eq. \ref{eq:newG}, demonstrating that both supersaturation with respect to ice and the ice water content over the course of the experiment are well-predicted by our new proposed model for depositional ice growth. The ability of our new proposed model for depositional ice growth learned from the levitation diffusion chamber experiments to reproduce ice growth in the AIDA Cloud Chamber indicates that this model generalizes well to realistic conditions for cirrus cloud evolution.

\section*{Conclusions and Outlook}
We have demonstrated the application of NODE's and symbolic regression to learn a mathematical model directly from multiple experimental time series of growing ice crystals. We validate our approach to learning the unknown part of a partially known ODE using a synthetic data set with a known functional dependence for the deposition coefficient $\alpha$. We find a weakly-constrained NODE model is able to more accurately reproduce the majority of our experimental time series for the growth of newly formed crystals, suggesting that prior parameterizations for surface kinetic effects as a single deposition coefficient function may be too restrictive. Indeed, this conclusion is consistent with \cite{pokrifka2020estimating} who found that newly formed crystals required transformations in the surface growth modes that control surface attachment kinetics.  We instead propose a new functional dependence for $G$ in Eq. \ref{eq:dmidt_simplified} that more accurately reproduces the experimental observations, and that can be expressed as a closed-form function using symbolic regression (Eq. \ref{eq:newG}). This function also generalized well to unseen data, both when predicting longer ice growth time series in the levitation diffusion chamber and when modeling observations of populations of ice crystals growing in the AIDA Cloud Chamber. 

This research contributes to the development of improved model parameterizations for cirrus cloud processes. Processes related to ice formation in clouds are a significant source of uncertainty in current climate models \cite[]{morrison2020confronting,duffy2024perturbing}. Detailed laboratory experiments are required to constrain individual microphysical process rates, but observables in laboratory experiments are often indirectly related to the prognostic variables used in atmospheric models.
Here we have focused on learning a single functional form that best fits all ice crystals; an alternative approach is to derive a distribution of growth rates from these experimental measurements \cite[]{pokrifka2023effective}. In addition to improving our fundamental understanding of the depositional ice growth process, the methods that we have introduced here can be more broadly applied towards parameterization-development of cloud microphysical processes \cite[]{lamb2025perspectives}. 

\section*{Materials and Methods}\label{sec:matmethods}
\subsection*{Depositional Ice Growth Model}\label{sec:proposedmodels}
The depositional ice growth rate for a single ice crystal growing from vapor (Figure \ref{fig:overviewmethods}a) in the atmosphere has traditionally been modeled using the capacitance ice growth model, which describes the growth of an ice crystal from the vapor phase \cite{pruppacher1998microphysics}. The capacitance growth model, including modifications for surface kinetic effects, can be written as an ordinary differential equation,
\begin{equation}\label{eq:dmidt}
\frac{dm}{dt} = 4\pi C (S_{i}-1)G(r,T,S_{i},\alpha),
\end{equation}
where $C$ is the capacitance, $S_{i}$ is the ice supersaturation far from the crystal, and $G$ represents the combined effects of vapor and thermal diffusivity to the surface of the ice crystal. $G$ is a function of temperature, pressure, and the modified diffusivity $D_{v}^{*}$. The capacitance is a function that depends on the geometry of the ice crystal; here we assume that $C = r$.

In the transitional regime between the continuum and the molecular limits, as is typically assumed for the early stages of atmospheric ice growth, the function $G$ given in \cite{LambVerlinde2011} is 
\begin{equation}\label{eq:g_sk}
G = \left[\frac{\mathcal{R}T_{\infty}}{e_{sat,i}(T_{\infty})D_{v}^{*}M_{w}} + \frac{L_{s}}{k_{a}T_{\infty}}\left(\frac{L_{s}M_{w}}{\mathcal{R}T_{\infty}}-1\right)\right]^{-1}.
\end{equation}

Due to surface attachment effects, the diffusivity in Eq. \ref{eq:g_sk} is modified from the continuum case. The modified diffusivity is defined below and includes influences of surface processes typically parameterized by a deposition coefficient $\alpha$. Surface attachment kinetics refer to the combined effects of individual vapor molecules adsorbing, migrating, and attaching to surface sites on a growing crystal. These collective processes lead to effective differences in vapor diffusivity relative to the continuum case (where surface attachment effects are ignored). Note that kinetic influences on thermal conduction are generally small and are ignored here \citep{harrington2019calculating}

Here we use the ``modified'' vapor diffusivity assuming spherical symmetry which can expressed as, 
\begin{equation}\label{eq:modifieddiff}
D_{v}^{*} = \frac{D_{v}}{\frac{r}{r+\Delta_{r}}+\frac{4D_{v}}{r\alpha w} } \mbox{  where  } w = \left(\frac{8RT_{a}}{\pi M_{w}}\right)^{1/2},
\end{equation}
where $D_{v}$ is the diffusivity of water molecules in air, $r$ is the radius of the ice crystal, $w$ is the molecular speed of water vapor in air, $M_{w}$ is the molecular weight of water, $R$ is the universal gas constant, and $T_{a}$ is the air temperature, $\Delta_{r}$ is the molecular jump distance, and $\alpha$ is the deposition coefficient. 

The deposition coefficient $\alpha$ in Eq. \ref{eq:modifieddiff} has previously been parameterized as a saturation and temperature dependent function in \cite[]{nelson1996new} as 
\begin{equation}\label{eq:AlphaNelson}
\alpha = \left(\frac{s_{local}}{s_{crit}}\right)^{m}tanh\left[\left(\frac{s_{crit}}{s_{local}}\right)^{m}\right].
\end{equation}
where $s_{local}$ is the supersaturation immediately above the crystal surface, $s_{crit}$ is the temperature-dependent critical supersaturation, and $m$ is a parameter which relates to the surface of the ice crystals. A value of $m=1$ is used to represent spiral dislocation growth, while a value of $m>10$ represents ledge nucleation. This parameterization is used to represent complex surface processes that are not fully represented by current theory. Here we use the temperature dependent parameterization for $s_{crit}$ from \cite[]{harrington2019calculating}. 
given by
\begin{equation}\label{eq:scrit}
s_{crit} = 9.6066 \times 10^{-5} T_{c}^{1.9171},
\end{equation}
where $T_{c}$ represents the temperature in Celsius.

In the continuum limit for a spherical ice crystal, $\alpha \rightarrow 1$ and Eq. \ref{eq:g_sk} reduces to
\begin{equation}\label{eq:Gcont}
G_c = G(D_v^*\rightarrow D_V).
\end{equation}

Eq. \ref{eq:dmidt} relies on a number of constants and empirical expressions, which we summarize in Supplementary Table S1 and Section S1. 
\subsection*{Levitation diffusion chamber data sets}\label{sec:experiments}

Experimental data was taken in the Button Electrode Levitation (BEL) thermal gradient diffusion chamber and were described in detail in \cite[]{harrison2016levitation,pokrifka2020estimating,pokrifka2023effective}. Charged droplets are initially levitated between two ice-coated parallel plates, with the bottom plate having an opposing direct current voltage, and the top plate an alternating current to stabilize particles horizontally. Due to differences in temperature between the warmer, upper plate and colder, lower plate, a supersaturation gradient exists in the chamber as a function of height. After the ice crystal nucleates, the voltage of the bottom plate is automatically adjusted to maintain constant levitation of the ice crystal, with the measured voltage being directly proportional to the ice crystal's mass as a function of time. The initial radius of the particle $r_{0}$ can be estimated from Mie theory using the diffraction pattern generated from the scattering of light from a Helium Neon laser, thus determining the initial crystal mass $m_{0}$. Deriving the mass as a function of time from the voltage measurement requires the assumption that ice is initially spherical, when it is likely poly-crystalline when it is initially nucleated \cite[]{bacon2003initial}. Since the uncertainty in derived mass is largely dominated by uncertainty in the initial size of the particle\cite[]{harrison2016levitation}, we optimize the NODE model against $m/m_{0}$ rather than $m$. 

Time series of the voltage of the lower plate are recorded at a 1 Hz frequency during the course of the experiments. Data sets consist of the measured temperature, pressure, supersaturation, initial ice particle radius observed from light scattering, and the voltage of the lower plate (proportional to the ice particle mass). Temperature uncertainty in the chamber is on the order of $<$1\% \cite[]{harrison2016levitation}. Supersaturation uncertainty is estimated to be on the order of 10\% \cite[]{pokrifka2020estimating}. Because supersaturation uncertainty is due to the location of the ice crystal in the chamber, it is likely to show a bias in one direction (typically low) although best estimates for the true value of the supersaturation are used in these data sets \cite[]{harrison2016levitation,pokrifka2020estimating,pokrifka2023effective}.

\subsection*{Synthetic data sets}\label{sec:synthetic}
To evaluate the method of using physics-informed machine learning to discover the functional dependence of ice growth from the mass ratios, we create synthetic data sets with a known depositional growth models. To simulate synthetic data, we start with the observed $S_{i}$, $T$, and $m_{0}$ from the 290 experiments that are shown in Supplementary Figure S1. Given these initial conditions for $m_{0}$, $S_{i}$, $T$, we assume depositional ice growth is described by Eq. \ref{eq:dmidt}, with the deposition coefficient function given by  Nelson and Baker \cite[]{nelson1996new} (Eq. \ref{eq:AlphaNelson}). We use an ODE solver to integrate Eq. \ref{eq:dmidt} to predict the evolution of the mass ratio for the same length as the experimentally measured time series (up to 500 s). Since the measured mass ratios in the levitation diffusion chamber have high frequency noise, we also use singular spectrum analysis (SSA) to determine realistic noise for the observed mass ratios. SSA decomposes time series into a sum of temporal principal components which account for a decreasing fraction of the variance in the original time series. Here we use a window size of 60 s and assume trailing eigenmodes represent the measurement uncertainty on $\frac{m}{m_{0}}$. An example of the simulated measurements for one time series is given in Supplementary Figure S3, showing the ODE solution and the SSA reconstruction from the trailing eigenmodes used to estimate measurement uncertainty. All 290 generated synthetic mass time series are shown in Supplementary Figure S2, which also shows the values for $\alpha$ predicted by Eq. \ref{eq:AlphaNelson} used in generating the synthetic data sets.  

\subsection*{Neural ODE model}
For both the strongly and weakly constrained models, we use a fully-connected neural network to parameterize the functional dependence in Eq.~\ref{eq:alphaNN} and Eq.~\ref{eq:GGcNN} with 3 linear layers with 50 neurons in each layer, and ReLU activation functions after the first and second layers. Following the third linear layer, we use a sigmoid activation function, as both functional forms are expected to be constrained to a range of values ($0 \leq \alpha\leq 1$ and $0 < \frac{G}{G_{c}} \leq 2$); without this constraint, the NODE model is significantly harder to optimize against observations. For the NODE, we tested both a fixed step (Runge-Kutta 4th order) and adaptive step-size (Runge-Kutta of order 8 of Dormand-Prince-Shampine) as implemented in the \texttt{torchdiffeq} library \cite[]{chen2018neural} to evaluate how this impacts the derived functional dependencies of Eq.~\ref{eq:alphaNN} and Eq.~\ref{eq:GGcNN}. In cases where the time series are less than 500 s, we mask experimental data such that it is not included in the calculation of the loss function. To minimize the loss function (Eq. \ref{eq:MSEloss}), we use the AdamW optimizer \cite[]{loshchilov2017decoupled}, with a base learning rate of 0.01, which we train for 500 epochs, with cosine decay \cite[]{loshchilov2016sgdr}. 

\subsection*{Symbolic Regression}
To find a symbolic expression for the function $\alpha = f_{\alpha}(S_{i},T|\theta_{\alpha})$ in terms of $S_{i}$ and $T$, we use the binary operators for summation, multiplication, exponentiation, division, and subtraction, and the unary operators $log$, $exp$, $1/x$, $square$, $cube$, and $tanh$, and run \texttt{PySR} for 1000 generations. 

To fit a symbolic expression to $G = G_{c}f_{G}(S_{i},T,m|\theta_{G})$, we use the binary operators for summation, multiplication, exponentiation, division, and subtraction, and the unary operators $1/x$, $square$, and $cube$, and run \texttt{PySR} for 1000 generations. We use $m$, $r$, $T$, $S_{i}$, and $G_{c}$ as input features, and target the prediction of $G$; this approach led to more accurate fits to the trained neural network than finding a symbolic expression for the ratio $G/G_{c}$. $G_{c}$ is given by Eq. \ref{eq:Gcont}, and does not depend on unknown or unobserved physics, such as the shape of the growing ice crystals.

\subsection*{AIDA chamber data sets}\label{sec:aidaexperiments}
To evaluate the performance of proposed depositional ice growth models on an independent set of observations, we compare against experiments of cirrus clouds in the AIDA Aerosol and Cloud Chamber. AIDA experiments were taken during the IsoCloud Campaigns, and simulated cirrus clouds between 195 - 235 K and pressures between 150 - 300 hPa, with both heterogenous and homogeneous ice nuclei. Average ice crystal sizes during these experiments are $<$ 10 $\mu m$ \cite[]{lamb2023re}. We use observations from the Chicago Water Isotope Spectrometer (ChiWIS)\cite[]{sarkozy_chicago_2020} of time-evolving vapor pressure of water vapor in the chamber, observations from the APeT instrument \cite[]{skrotzki2012high,ebert2008simultaneous} measuring after a heated inlet to observe the total water vapor (ice and water vapor) inside the chamber, and measurements of the number concentrations of ice crystals from the welas optical particle counters. In addition, the temperature and pressure of the gas inside the chamber is monitored during experiments. Additional details of the measurements and experiments are given in \cite[]{lamb2017laboratory,clouser_no_2020,lamb2023re}. To compare against the experimentally observed IWC, we use a bin microphysics model based on the one described in \cite[]{lamb2023re}.

\subsection*{Data, Materials, and Software Availability}
Code to reproduce the NODE models, training process, and symbolic regression, and the comparison against the AIDA experiments is provided in the github repository \href{https://github.com/kdlamb/IceNODE}{https://github.com/kdlamb/IceNODE}. Data sets for the levitation diffusion chamber experiments used in this analysis are available at doi:10.26208/dd1w-wa17, doi:10.26208/z7bf-nq20, doi:10.26208/htw5-q166. Data sets for the IsoCloud 4 campaigns can be found at \href{https://zenodo.org/records/7986868}{https://zenodo.org/records/7986868}. An earlier version of this paper was presented at the 2024 Conference on Neural Information Processing Systems Machine Learning and the Physical Sciences Workshop.

\subsection*{Acknowledgements}
We acknowledge support from the Department of Energy (DOE) Atmospheric Science Research (ASR) grant DE-SC0023020 and from the NSF through the Learning the Earth with Artificial Intelligence and Physics (LEAP) Science and Technology Center (STC) (Award \#2019625). J.Y.H. acknowledges the support of NSF AGS-2128347 and NSF AGS-2516531 for the measurements. We also acknowledge Elisabeth J. Moyer, Laszlo Sarkozy, and the IsoCloud science team and the AIDA technical staff and support team. Funding for the IsoCloud campaign was provided by the National Science Foundation (NSF) and the Deutsche Forschungsgemeinschaft through International Collaboration in Chemistry grants CHEM1026830 and MO 668/3-1.

\onecolumn

\bibliography{IceSciMLreferences}

\clearpage
\section*{Supplementary Information}

\setcounter{table}{0}
\setcounter{figure}{0}

\renewcommand{\thetable}{S\arabic{table}}
\renewcommand{\thefigure}{S\arabic{figure}}

Here we provide addition figures and tables in support of the analysis in the main document.

\subsection*{Constants and Parameterizations used in the Depositional Ice Growth Model}
For the depositional ice growth model, we assume that the temperature dependence of saturation vapor pressure with respect to ice (in Pa) is given by \cite[]{murphy2005review}, 
\begin{equation}
    e_{sat,i}(T) = exp\left(a_{0}-\frac{a_{1}}{T}+a_{2}Log(T)-a_{3}T\right)
\end{equation}
where $a_0=9.550426$, $a_{1}=5723.265$, $a_{2}=3.53068$, and $a_{3}=0.00728332$.

The diffusivity of water vapor in air (in $m^{2}/s$) is given by
\begin{equation}
    D_{v} = 2.11\times10^{-5}\left(\frac{T}{T_{0}}\right)^{1.94}\left(\frac{p_{0}}{p}\right)
\end{equation}
where $T_{0}=273.15$ K and $p_{0}=101325$ Pa \cite[]{pruppacher1998microphysics}, which is valid for temperatures between -40 to 40 $^{\circ}C$.

The thermal conductivity of air in Joules is given by
\begin{equation}
k_{a} = 4.187\times10^{-3}(5.69+0.017(T-273.15))
\end{equation}
Additional constants are given in Table \ref{table:constants}.

\vspace{1cm}
\begin{table}[h!]
\caption{Values of constants used in depositional ice growth model.\label{table:constants}}
\centering
\begin{tabular}{lcc}
\hline
Name & Symbol & Value   \\
\hline
Density of water & $\rho_{w}$ & 1000 kg/m$^{3}$   \\
Density of ice & $\rho_{i}$ & 910 kg/m$^{3}$   \\
Molecular mass of water & $M_{w}$ & 18 $\times 10^{-3}$ kg   \\
Thermal deposition coefficient & $\alpha_{T}$ & 1  \\
Latent heat of vaporization & $L_{v}$ & 2.5$\times 10^{6}$ J/kg \\
Latent heat of sublimation & $L_{s}$ & 2.837$\times 10^{6}$ J/kg\\
Acceleration due to gravity & $g$ & 9.81 $m/s^{2}$ \\

Universal gas constant & $\mathcal{R}$ & 8.3144521 J/mole/K  \\
Individual gas constant of air & R$_{a}$ & 287.05 J/kg/K  \\
Individual gas constant of water vapor & R$_{v}$ & 461.51 J/kg/K  \\

Specific heat capacity  & c$_{p}$ & 1005   \\
Mean free path of water molecules in air & $\lambda_{a}$ & $8\times 10^{-8}$ m\\
Vapor jump length & $\Delta_{v}$ & 1.3 $\lambda_{a}$\\
Thermal jump length & $\Delta_{T}$ & $2.16\times 10^{-7}$ m\\
\hline
\end{tabular}
\end{table}

\newpage

\begin{figure}[ht!]
\noindent\includegraphics[width=\textwidth]{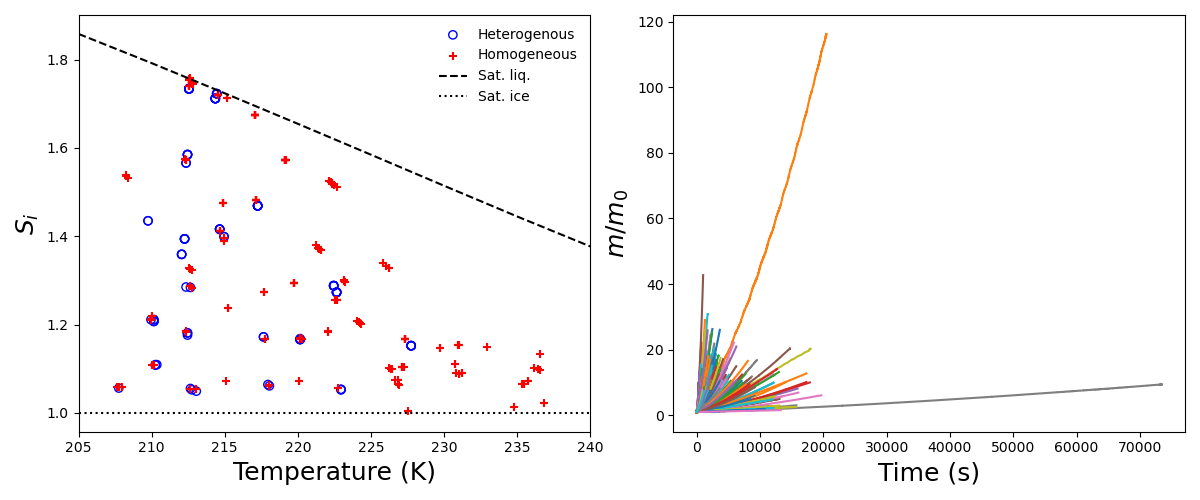}
\caption{\textbf{Overview of experimental data sets used in this analysis.} \textbf{Left:} Saturation with respect to ice and temperature at which experiments included in this analysis were performed. The dashed line shows the temperature dependence of saturation with respect to liquid water, while the dotted line show saturation with respect to ice. Symbols indicate whether ice crystals were nucleated heterogeneously or homogeneously. \textbf{Right:} Mass ratios of all ice crystals as a function of time.}
\label{fig:expoverview}
\end{figure}

\begin{figure}[ht!]
\noindent\includegraphics[width=\textwidth]{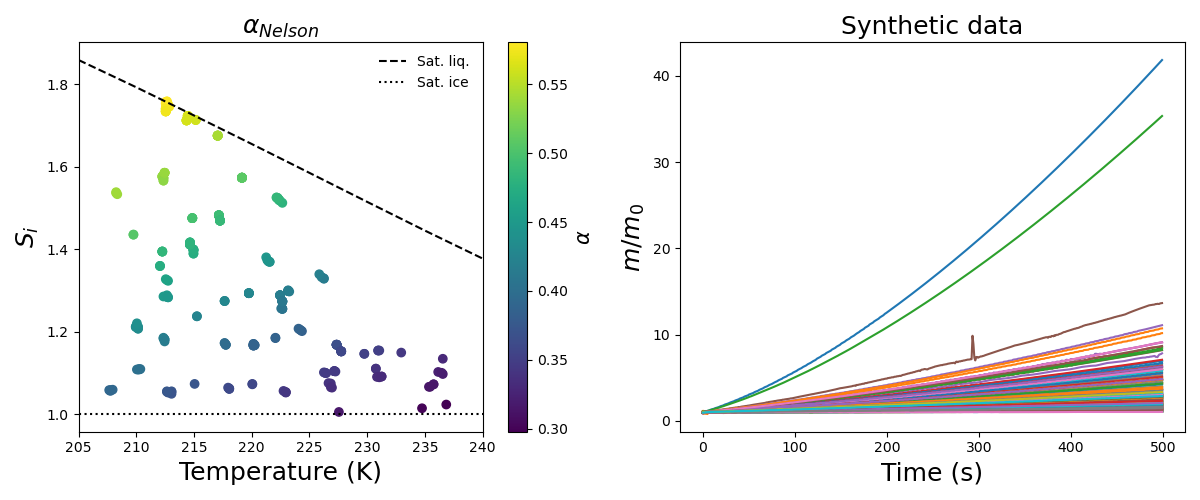}
\caption{\textbf{Synthetic data sets.} \textbf{Left:} Values for $\alpha$ predicted by Nelson and Baker, 1996 \cite[]{nelson1996new} parameterization for the $S_{i}$ and T conditions for the 290 experimental data sets (used as initial conditions in creating the synthetic mass time series). \textbf{Right:} Times series of mass ratios for all of the synthetic data. }
\label{fig:synthoverview}
\end{figure}

\begin{figure}[ht!]
\begin{center}
\noindent\includegraphics[width=0.45\textwidth]{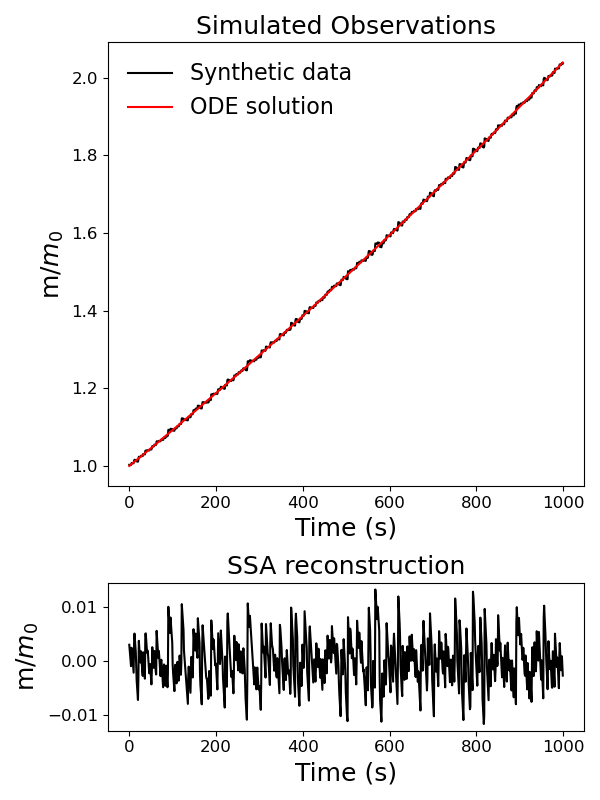}
\caption{Example of synthetic time series for mass ratios, with measurement uncertainty derived from trailing eigenmodes of singular spectrum analysis of observed time series.}
\label{fig:syntheticMratio}
\end{center}
\end{figure}

\begin{figure}[ht!]
\noindent\includegraphics[width=\textwidth]{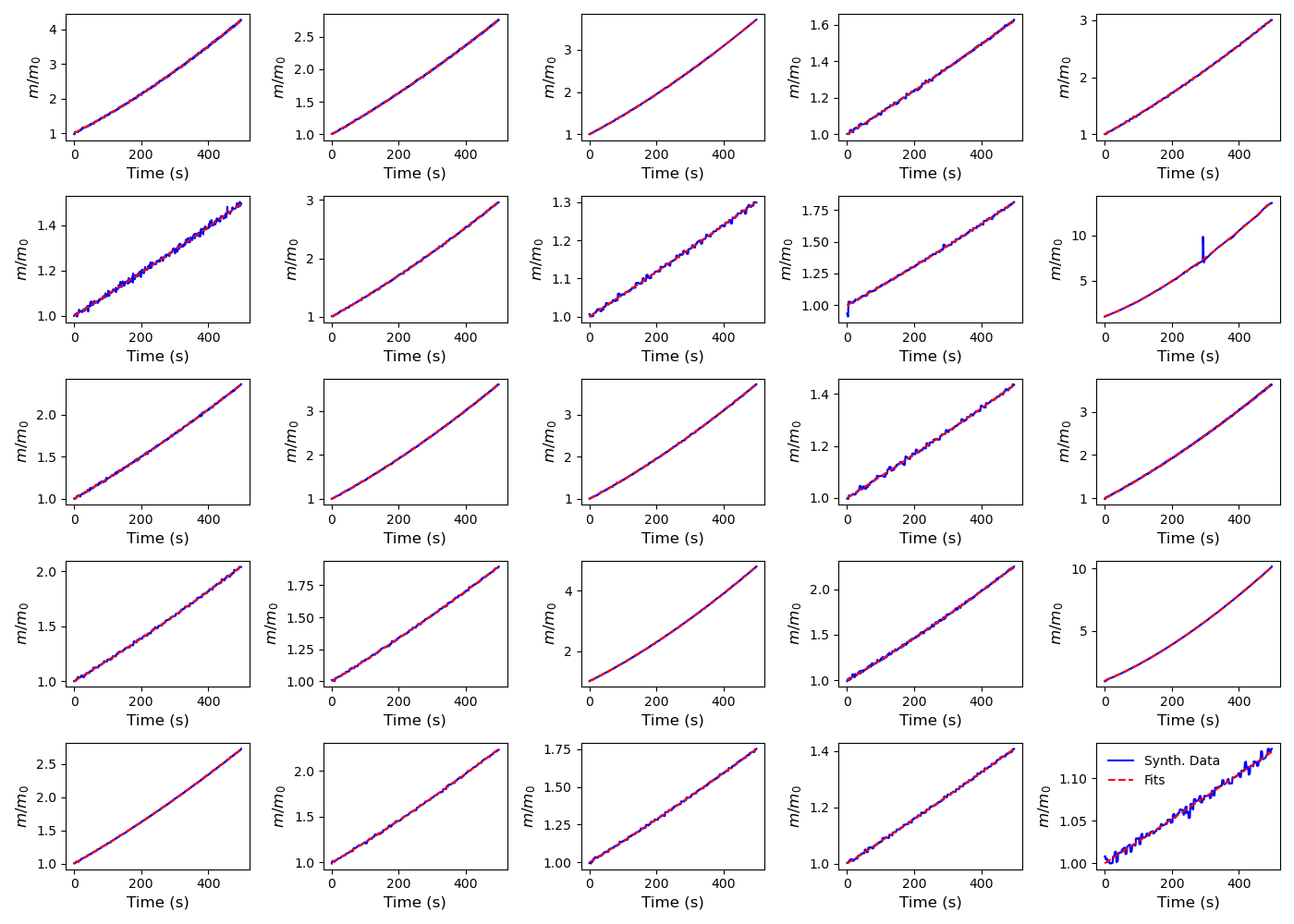}
\caption{\textbf{Model predictions for individual experiments for the synthetic data sets.} Synthetic mass ratios and strongly-constrained NODE model predictions for a subset of the time series.}
\label{fig:synthetictimeseries}
\end{figure}

\begin{figure}[ht!]
\noindent\includegraphics[width=\textwidth]{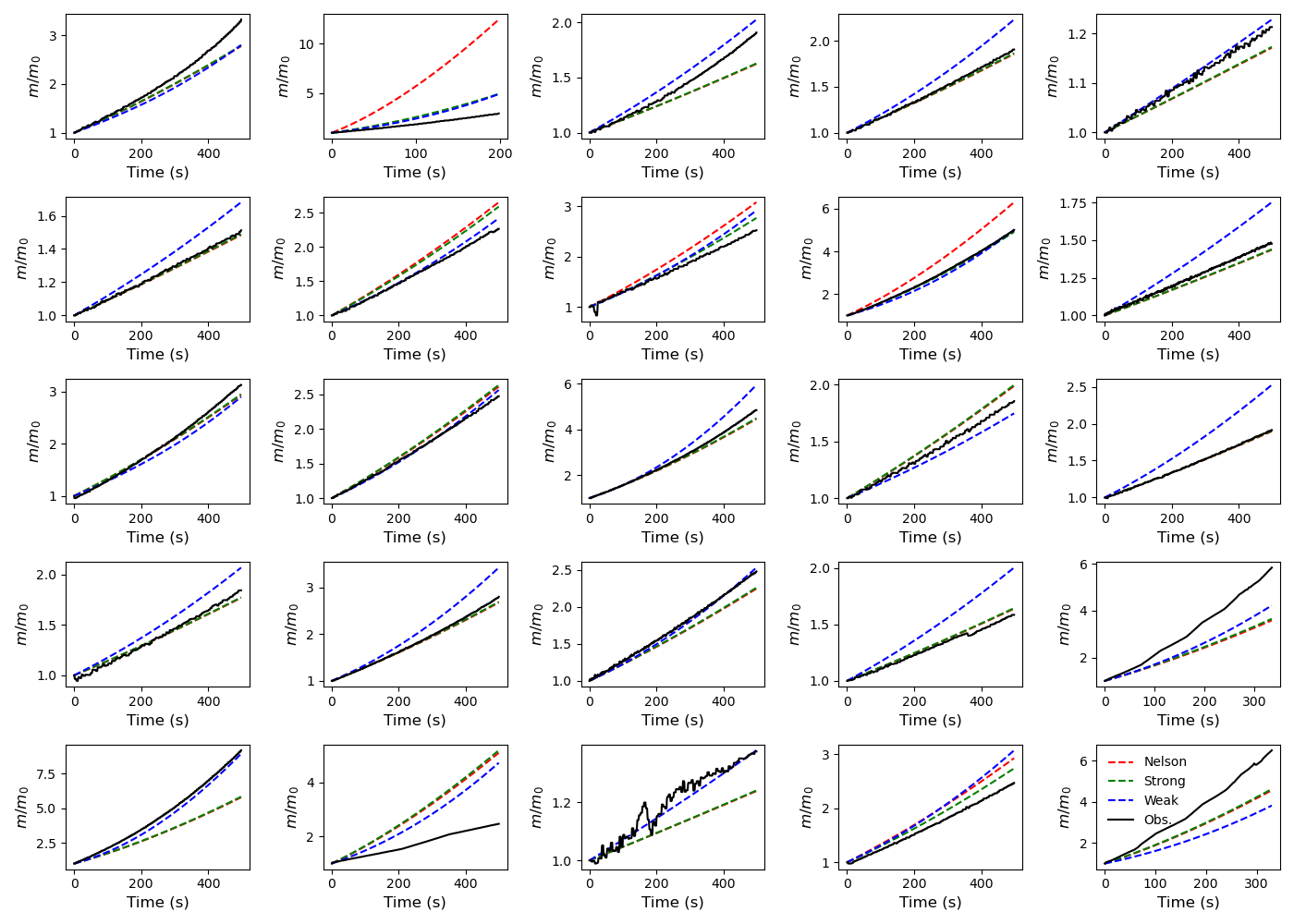}
\caption{\textbf{Model predictions for individual experiments for the real data sets.} Nelson and Baker model predictions, strongly-constrained NODE model predictions, and weakly-constrained NODE model predictions compared to observations for a subset of the time series.}
\label{fig:realtimeseries}
\end{figure}

\begin{figure}[ht!]
\begin{center}
\noindent\includegraphics[width=1.0\textwidth]{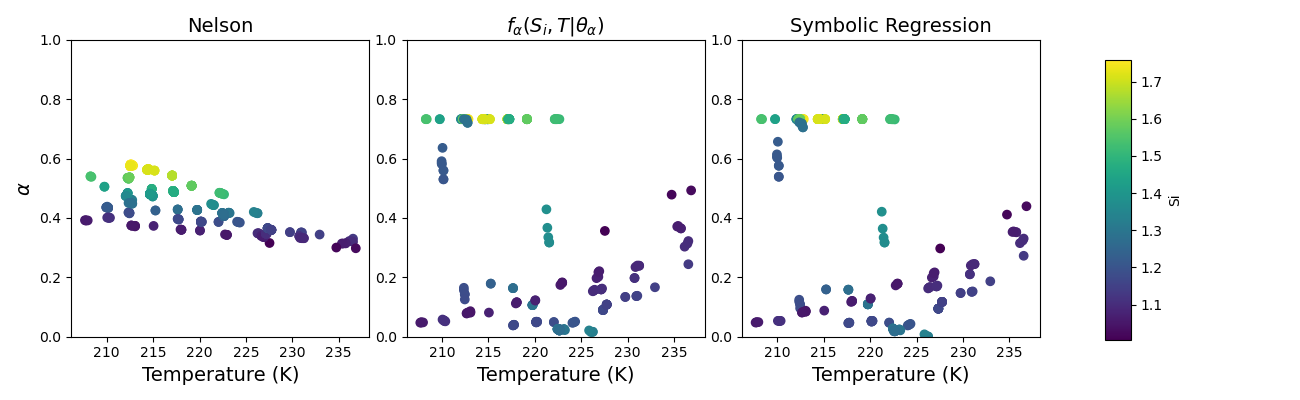}
\caption{\textbf{Functional dependence of $\alpha$ learned from the real data.} a) Saturation and temperature dependence of $\alpha$ parameterization from Nelson and Baker, 1996 \cite[]{nelson1996new}. b) Saturation and temperature dependence from the trained NN for the real data sets. c) Predictions from an expression learned by symbolic regression from the trained NN for the real data sets.}
\label{fig:realstrongalpha}
\end{center}
\end{figure}

\newpage

\begin{table}
\def\arraystretch{2.5}
\caption{Equations for $G$ discovered by symbolic regression.\label{table:equations}}
\centering

\begin{tabular}{|l|l|}
\hline
\textbf{\#} & \textbf{Discovered Equation}\\
\hline
0 & $G_{c}$\\
\hline
1 & $0.93458 \times G_{c}$\\
\hline
2 & $G_{c} - 0.347\times10^{-21} m^{-1}$\\
\hline
3 & $652.8 \times G_{c} m^{0.253}$\\
\hline
4 & $\frac{G_{c}}{3.50179\times10^{-6} m^{-0.469} + 0.413}$\\
\hline
5 & $\frac{G_{c}}{2.89192\times 10^{-6} m^{-0.476} + 0.419}$\\
\hline
6 & $G_{c}\left[0.615 + \frac{3.13\times10^{-9}}{0.836 G_{c} + 1000m}\right]^{-1}$\\
\hline
7 & $G_{c}\left[0.615 + \frac{3.13\times10^{-9}}{0.902 G_{c} + 1000 m}\right]^{-1}$\\
\hline
8 & $688.267 G_{c}^{1.3153}\left[0.85601 + \frac{2.6606\times10^{-12}}{m}\right]^{-1}+0.1123\times10^{-9}$\\
\hline
9 & $\left[\left(0.803 + 2.16\times10^{-12}m^{-1}\right) \left(2.7649\times10^{-7}G_{c}^{-0.743} - 0.360\right)\right]^{-1}\times10^{-9}$\\
\hline
10 & $\left[{\left(0.504 + \frac{2.86}{m\times10^{12} +2.08}\right) \left(1.2287^{-7}G_{c}^{-0.795} - 0.390\right)}\right]^{-1}\times10^{-9}$\\
\hline
11 & $\left[\left(1.1805\times10^{-7}G_{c}^{-0.797} - 0.389\right) \left(0.5018 + \frac{2.85}{m\times 10^{12} + 2.06}\right)\right]^{-1}\times10^{-9}$\\
\hline
12 & $\left[\left( 0.591 + \frac{2.83}{\frac{m\times10^{12} +0.726}{0.736} + 0.224}\right) \left(\left(\frac{2.01}{G_{c}\times10^{9}}\right)^{0.799} - 0.409\right) - 0.355 + \frac{0.340}{\left(\frac{1.61}{G_{c}\times10^{9}}\right)^{(S_{i}-1)}}\right]^{-1}\times10^{-9}$\\

\hline
\end{tabular}
\end{table}

\begin{figure}[ht!]
\centering
\noindent\includegraphics[width=1.0\textwidth]{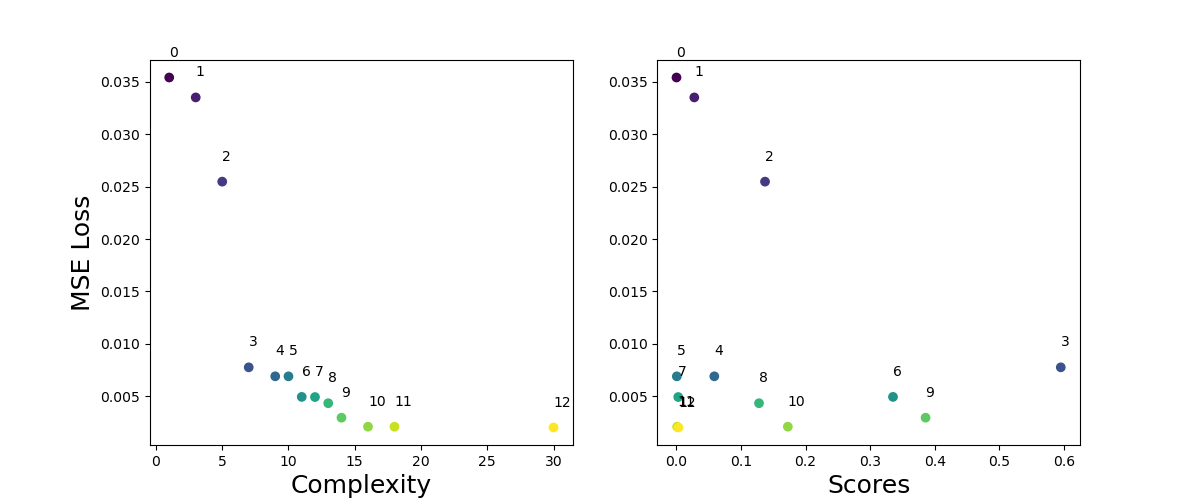}
\caption{\textbf{Comparison of MSE loss (after 500 seconds) for each symbolic regression expression.}  Equation numbers correspond to the rows in Table \ref{table:equations}. The complexity of the equation is defined as the number of a nodes in an expression tree. The score is defined as the negative of the log-loss with respect to complexity.}
\label{fig:bestexpSRcomparisons}
\end{figure}

\begin{figure}[ht!]
\centering
\noindent\includegraphics[width=0.7\textwidth]{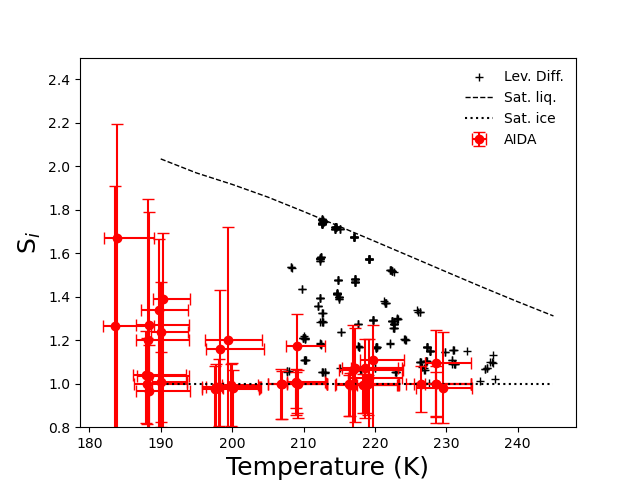}
\caption{\textbf{Temperature and Supersaturation range for AIDA experiments compared with levitation diffusion chamber.} }
\label{fig:aidarange}
\end{figure}

\begin{figure}[ht!]
\centering
\noindent\includegraphics[width=1.0\textwidth]{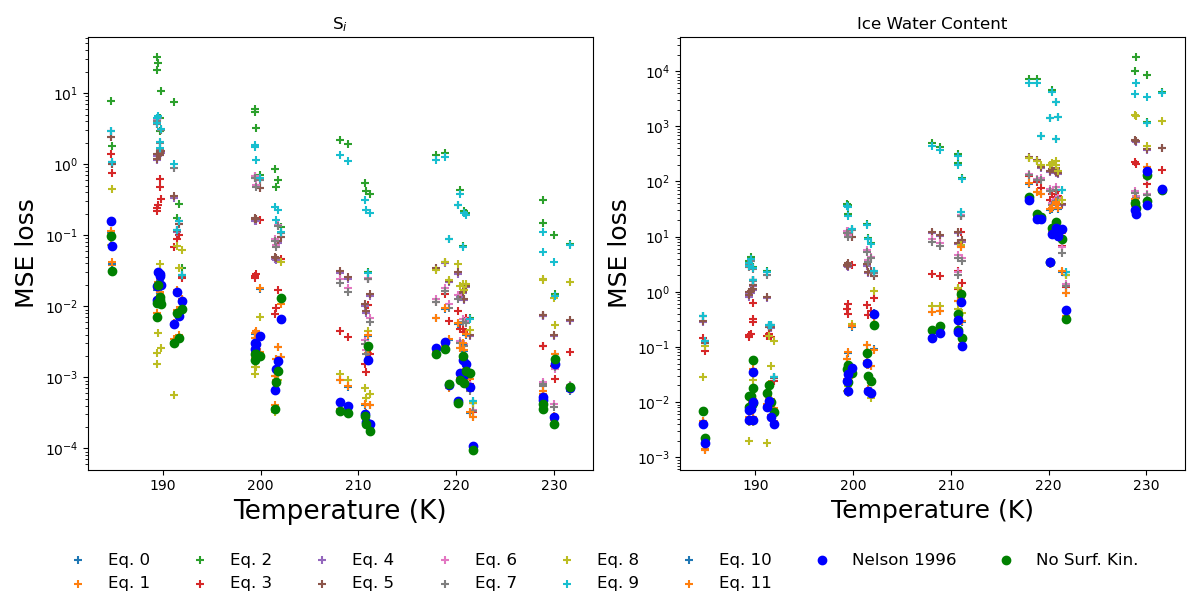}
\caption{\textbf{Comparison between AIDA experiments and bin microphysics model}. MSE loss for $S_{i}$ (left) and IWC (right) between bin microphysics model using discovered equations for G given in Table S2 and observations from AIDA.}
\label{fig:aidaexps}
\end{figure}

\end{document}